# Interaction of Molecular Oxygen with Hexagonally Reconstructed Au (001) Surface

Andrew Loheac[1], Andi Barbour[2*], Vladimir Komanicky[3], Chenhui Zhu[2†], John Collini[1], Anthony Ruffino[1], Yihua Liu[2], Hoydoo You[2] and Michael S. Pierce[1]

[1]Rochester Institute of Technology, Department of Physics, Rochester NY, 14623
[2]Argonne National Laboratory, Materials Science Division, Argonne IL 60439
[3]Safarik University, Faculty of Science, Košice, Slovakia

Kinetics of molecular oxygen / Au (001) surface interaction has been studied at high temperature and near atmospheric pressures of $O_2$ gas with *in situ* x-ray scattering measurements. We find that the hexagonal reconstruction (hex) of Au (001) surface lifts to (1×1) in the presence of $O_2$ gas, indicating that the (1×1) is more favored when some oxygen atoms present on the surface. The measured lifting rate constant vs. temperature is found to be highest at intermediate temperature exhibiting a 'volcano'-type behavior. At low temperature, the hex-to-(1×1) activation barrier ($E_{act}$ = 1.3(3) eV) limits the lifting. At high temperature, oxygen adsorption energy ($E_{ads}$ = 1.6(2) eV) limits the lifting. The (1×1)-to-hex activation barrier ($E_{hex}$ = 0.41(14) eV) is also obtained from hex recovery kinetics. The pressure–temperature (PT) surface phase diagram obtained in this study shows three regions: hex at low P and T, (1×1) at high P and T, and coexistence of the hex and (1×1) at the intermediate P and T.

---

[*] Current Address: Brookhaven National Laboratory, Upton NY 11973
[†] Current Address: Lawrence Berkeley National Laboratory, Berkeley CA 94720

The study of catalysis on model surfaces can provide important benchmarks that expand our understanding of chemical processes and improve real-world application. Understanding the physical characteristics of material surfaces at an atomic and molecular level, and how they relate to chemical activity, is a driving factor in the development of modern catalysts. However, while a significant amount of surface science has been successfully conducted under ultra-high vacuum conditions, these idealized environments often do not fully reflect the behavior of material surfaces in atmospheric and oxidizing conditions under which most catalysts operate. The surface of elemental Au is one such model surface and studies of the atomic scale interactions at high pressure and temperature are limited.

In most common, ambient conditions Au is relatively inert and unreactive, due in large part to a filling of the antibonding d-states [1], which includes a filled 5d shell with a shielded 6s shell [2]. However, there are preparation conditions where the Au shows chemical activity [3]. In some instances there is significant interaction between oxygen and Au surfaces, even for the formation of chemisorbed oxygen for both extended Au (111) facets [4, 5, 6, 7, 8] and nano-particle prepared on oxides [9, 10, 11]. The interaction of oxygen with Au also presents interesting challenges for density functional theory (DFT) calculations [12, 13, 14].

In general, a dissociation process for molecular oxygen, such as dosing over a hot filament [15, 16], ozone decomposition [17, 18], or electron-induced dissociation [19], is required before any significant interaction can occur at gold surfaces. In the case of high temperatures, above 700-800 K, it is possible for oxygen to chemisorb on the Au (111) surface directly from dioxygen gas [7, 8, 20]. More recently it was shown that the surfaces of supported gold nano-particles can achieve this at much lower temperatures [10]. While the stability of such oxygen on the Au (111) surface can depend upon the presence of contaminants [4, 15], the formation of an Au-oxygen surface layer on a clean surface at high pressures and temperatures is well established [5, 6, 7].

The Au (001) surface exhibits a well-known quasi-hexagonal (hex) surface reconstruction [21] that continues to exhibit interesting character. The corrugated top hexagonal layer is slightly rotated (<1°) away from the main cubic axes and x-ray scattering has provided the precise structure [22, 23, 24, 25] in vacuum and in electrolytes [26, 27]. In fact, interaction of *atomic* oxygen with the Au (001) surface has been investigated before using filament dosing [16] and temperature programmed desorption of sputtered oxygen [28], both revealing the adsorption of atomic oxygen capable of lifting the hex reconstruction. However, interaction of *molecular* oxygen has not been studied so far to our knowledge.

**Experimental Details**

The experiments were performed at beamline 33-BM of the Advanced Photon Source at Argonne National Laboratory in surface scattering geometry. Au single crystals, 6mm in diameter and 4-5mm tall, were cut and polished on the (001) facet. For these experiments, the Au single crystals were mounted on quartz pedestals and grounded to prevent surface charging during x-ray exposure. The samples were placed in a quartz cell and that mounted directly to the 4-circle diffractometer, allowing access to in-plane scattering vectors. Mass flow controllers maintained relative concentrations between 0.5 and 100% flowing over the sample surface at atmospheric pressures. He gas, balanced with 2% $H_2$, was used to provide a neutral (oxygen-free) environment during the experiments, flowing at a rate of 300 sccm. Experiments were also performed without $H_2$, but these did not show significant variation from those with $H_2$. The photon energy was set to 17.0 keV and the beam focused to 0.2×0.1 mm at the sample position. The resulting surface scattering provided a signal of ~ $10^4$ x-rays per second. A Pilatus 100K area detector provided efficient data collection, as well as eliminating most of the Au L-edge fluorescence. Temperature was controlled by a radio frequency induction heating coil outside the cell. The RF heater coupled only directly to the metal crystal leaving only conductive heating to the inside of the quartz. The temperature of the crystal was measured during the experiments by tracking the bulk lattice parameter. An example of the sample and schematic of the cell are shown in Fig. 1, along with locations of the surface diffraction peaks and an example x-ray CCD image.

The samples and cell components were cleaned by boiling in nitric acid for 20 minutes. The samples were then annealed at ~ 95% of the bulk melting point for 24 hours, resulting in a mosaic spread of 0.15 degrees or less. The samples were then stored in ultrapure deionized water with resistivity of 18 MΩ•cm and transported to the beamline. Once installed, the sample surface was prepared by annealing the crystal to ~90 % of the bulk melting point. This served as a reliable method of restoring the surface reconstruction. The resulting surface signal was similar to other experiments, both those conducted in ultra-high vacuum (UHV) and ambient conditions. The sample, when prepared in-situ and cooled in the presence of the inert gas environment exhibits the well known hex surface reconstruction, along with the presence of the low temperature rotation domains as shown in Fig. 2.

Crystal Truncation Rod (CTR) measurements are a proven technique for determining the average local structure of atoms at an interface [29]. Following changes to the diffracted intensity at key positions along a CTR is an established technique to measure time-resolved changes to the surface [30]. This method has recently been used to study the interaction of CO with the Au (001) surface [31]. The hex reconstruction for Au (001) produces peaks at (1.2, 1.2) in the (H,K) surface plane, and at 30 degree rotations from that position as indicated in the upper right diagram in Fig. 1b. In these experiments (1.2, 1.2, L) was used to access the surface information with L being 0.1 or 0.15 reciprocal lattice units. Offline image analysis was used to sum the peak intensity and remove the background parasitic scattering. This was sufficient to remove the effects of the remaining fluorescence as well as the development of an additional small contribution to the background due to Debye-Waller effects at high temperature.

**Structure in the presence of Oxygen**

Examples of the intensity across the hex peak position are shown in Fig. 2 for different conditions. For low temperatures the rotated hex phase commonly seen in vacuum studies is present as indicated by two satellite peaks situated at about ± 1 degree apart from the center peak; however, at the temperatures typical to this study (T > 900 K), the rotation of the hex phase is absent. Instead the surface hex aligns to the main axes of underlying bulk cubic lattice. Once oxygen is introduced at sufficient pressure, the observed hex intensity decreases. While our extensive search near ambient atmospheric $O_2$ pressure detected no ordered surface oxide formation, we find a significant change in the hex reconstruction even under a minute amount of $O_2$ exposure. If the increase in the oxygen partial pressure or temperature is sufficient, the hex phase will vanish entirely even in the absence of x-rays. Remarkably, the decrease in intensity, regardless of pressure, accompanies an ~ 0.1% expansion of the hexagonal lattice. We believe such an expansion is consistent with the hex layer expanding to accommodate oxygen atoms adsorbed on the hexagonal lattice even though we neither directly detect oxygen over-layers nor determine the adsorption sites. Note that the entirety of the hex domain would expand quickly within seconds upon dosing $O_2$ to the chamber, while the hex intensity would continue to decrease over minutes. This indicates that the hex phase is significantly more active in $O_2$ reduction than the Au (111) surface where $O_2$ is expected to interact too weakly to reduce [32].

CTR measurements were collected along the (00L) and (11L) directions in the presence and absence of oxygen. The intensity variation was modeled using an adaptation of the Au (001) surface in vacuum [24]. Fits using this model are shown in Fig. 3. The fit lines are constrained to only allow variation in the surface parameters, and are consistent with a hex reconstruction in the absence of oxygen and a disordered (1 x 1) surface in the presence of oxygen. In the absence of oxygen, the density of the top layer is 1.25 (.05) relative to a bulk Au layer. While in the presence of 2.5% atm oxygen at 1150 K the density of the top layer is 0.21 (.04) that of the bulk layers. Both surface layers were found to be relaxed by ~ 20% of the bulk lattice spacing. CTR modeling including the presence of significant oxygen in the top layer was incapable of improving the fitting in any statistically meaningful fashion. Thus the CTR data is consistent with a lifting of the surface reconstruction and a disordered bulk truncation in the presence of oxygen. This lack of ordered structure is consistent with earlier observations of atomic oxygen dosing of the Au (001) surface at lower temperatures [16].

**Partial Pressure – Temperature surface phase diagram**

A surface pressure-temperature (PT) phase diagram was created from the measurements of the equilibrium intensity of the hex peak. Since the intensity of the hex peak is a direct measure of the surface state, it was possible to directly observe the effect of oxygen on the surface. Full intensity indicates an unrotated hex reconstructed surface, and an absence of intensity corresponds to the disordered (1×1) surface state. The hex peak would not fully lift at many of the lower oxygen concentrations. Instead the intensity would become constant at an intermediate value indicating phase coexistence.

The pressure-temperature surface phase diagram is shown in Fig. 5 with the regions indicated. The red squares indicate regions where only disordered (1×1) exists, defined by no observable hex intensity above the background signal, typically corresponding to the highest temperatures and pressures. We define the hex region, shown by the blue circles, by the intensity drop less than 5% upon introduction of oxygen. The coexistence region indicated by the green diamonds is between these limits. The boundaries on the phase diagram are guides to the eye.

Vacuum studies have found the temperature of the hex reconstruction to lift at 1170 K [24, 25, 26]. This lifting temperature agrees well with the lifting temperature in the absence of oxygen. The phase diagram indicates that oxygen destabilizes the surface reconstruction and lifts the hex at lower temperature. The adsorbed oxygen reduces the free energy of the (1x1) state such that it becomes favorable at some temperatures below 1170 K.

Several isobar intensities are shown in Fig. 4. For significant oxygen partial pressures, such as 2.5% and 3.3% shown, the hex intensity slowly dropped by 10-20% over several hundred degrees, before rapidly decreasing to zero. At such high oxygen pressures, a phase coexistence region exists between ~900 K and 1130 K. At lower oxygen concentrations the hex remains largely, if not fully, reconstructed over these intermediate temperatures. For lower oxygen pressures, partial lifting is observed only in the narrow temperature range of ~ 10s of degrees below 1170 K.

It is possible to model the change in intensity with temperature. With the assumption that the intensity measures the fraction of surface in the hex phase, and that in equilibrium the coverage of the surface is proportional to a Boltzmann factor, the temperature and pressure dependence of the hex fraction can be expressed with a functional form of

$$\text{Eq. (1): } \theta_{hex} = 1 - e^{-\alpha/p_{O_2}(1-T/T_L)}$$

In this expression $p_{O_2}$ represents the fractional pressure of the oxygen and $T_L$ is the temperature at which the reconstruction appears to be totally lifted as a function of pressure. Remarkably, this can be further parameterized as $T_L = T_C(1 - \gamma p_{O_2})$, where $T_C$ is the temperature at which the surface reconstruction lifts in vacuum and $\gamma p_{O_2}$ gives the reduction in temperature for lifting. Taking the measured value of $T_C$ from literature value of 1170 K [24, 25], this expression is capable of producing the fit lines shown in Fig. 4 with only two adjustable parameters. However, the constants $\alpha$ and $\gamma$ have no established relationship to thermodynamic parameters.

Once the non-isobar data points are included, a surface phase diagram can easily be generated as shown in Fig. 5. The phase diagram can be compared to the CO-induced lifting [31] of the Au (001) reconstruction. The hex lifts at low CO partial pressure for low temperatures and requires higher CO pressure for higher Au temperatures. The phase boundary follows a constant chemical potential line indicating that CO adsorption directly lifts the hex reconstruction [31]. In the current study with $O_2$, however, we can see that the increased partial pressure *lowers* the lifting temperature as shown in Fig. 5(a). This suggests that $O_2$ adsorption lifts hex indirectly. Since $O_2$ cannot adsorb directly to the surface, $O_2$ must be first physisorbed and dissociate (reduction) for atomic adsorption. The dissociated oxygen atoms may favor non-hex sites such as step edges, defects, and the (1×1) sites due to significantly higher adsorption energy compared to the regular triangular lattices [33]. As a result, the (1×1) structure with oxygen adsorbed becomes lower in energy than the hex, which leads to lifting of hex. We believe this indirect lifting, which will be the subject of the kinetic measurements, is the reason for the phase boundary with the temperature dependence opposite to the case of CO adsorption.

**Kinetics between hex and (1x1)**

The intensity of the hex peak the surface transition could be observed in real-time as shown in Fig. 6. Furthermore, the observed rate of change of the x-ray intensity in response to oxygen exposure empirically follows a simple first order kinetics reaction. As shown, upon exposure to oxygen the intensity would decay following a simple exponential $I(t) \propto e^{-t/\tau}$ and upon the removal of oxygen the intensity would recover as $I(t) \propto (1 - e^{-t/\tau})$. This behavior was explored over the data points that make up the co-existence region of the phase diagram. It reveals that over a significant range of pressures and temperatures the lifting of the hex reconstruction can be modeled by a first order chemical process as demonstrated by the fit lines in Fig. 6. For relatively low pressures, those insufficient to completely lift the hex reconstruction, the rate of lifting shows little dependence upon pressure, instead only significantly varying with changes in temperature. Evolution of the surface in the presence and absence of x-rays produced identical results, eliminating the x-rays as a contributing factor to the observed kinetics. Further details are given in the supplemental information.

The hex to (1×1) transition rate as a function of temperature is shown in Fig. 7a. The rate is observed to first increase with temperature, followed by an eventual decrease for T > 950 K. This rise and subsequent decrease in the rate of lifting at higher temperature cannot be described by a single simple Arrhenius relation, indicating operation of two competing processes. It is plausible to assume that the competing processes are the hex-to-(1×1) activation process dominating at low temperature and the loss of oxygen coverage in the (1×1) structure at high temperature as hypothesized for the phase diagram. In Fig. 7b, an energy level diagram showing schematically that the presence of atomic oxygen on the surface lowers the free energy of the (1×1) structure. Once $O_2$ is in contact with the high temperature surface, a small fraction adsorbs dissociatively and the free energy diagram changes from the blue curve to the red curve. If there is sufficient thermal energy to overcome the energy barrier $E_L$, then the reconstruction lifts to form a disordered (1×1) state. In effect, the presence of the oxygen causes the (1×1) state to be temporarily stabilized. At lower temperatures there is insufficient energy for this process to occur and no direct interaction between molecular oxygen and the Au surface can occur. At high temperature, the oxygen will also readily leave the (1×1) surface, allowing the reconstruction to reform. The oxygen coverage will depend on the oxygen pressure allowing the (1×1) energy to vary between these two energy levels ($H_{Ads}$).

It is most likely that the lifting process begins at step edges, defects, or boundaries of existing (1×1). As such that the stability of the Au hex is locally compromised. It is expected that the energy difference between Au hex + $O_2$ (gas) and the adsorbed state Au hex + O is relatively small due to the absence of a stable Au-oxide. The relative energy difference of these two states shown in the Fig. 7b diagram are purposefully small to indicate this, however is not directly measured. Under this scenario, the kinetics at low temperatures are slow due to an activation barrier, yet also slow at high temperature because of high desorption rate of oxygen from the non-hex (1×1) sites. This results in a volcano-type activity where the activity is highest for intermediate temperatures.

It is possible to clarify this more explicitly. For equilibrium populations, the rate equation largely depends upon the final state energetics of the phases [34]. Given that these measurements directly probe the ordering of the Au atoms at the interface, the rate equation for the lifting of the surface reconstruction can be written as

$$\text{Eq. (2): } \frac{d\theta_{hex}}{dt} = -A\theta_{hex} e^{-E_L/k_B T}$$

Here $\theta_{hex}$ represents the hex coverage, $E_L$ the activation barrier shown in Fig. 7b, and a proportionality constant $A$. The activation barrier is approximately the energy required for a surface Au atom jump out from its surface site to on-top position, which is required for decreasing density of the surface layer during the lifting. Both adsorption and desorption of the oxygen are occurring and the oxygen coverage will depend upon temperature. If it is assumed that the process is dissociative, there will be a $(1-\theta_O)^2$ dependence. With the heat of adsorption, $H_{Ads}$, shown in Fig. 7b, and $B$ as an additional prefactor containing available sites and frequency components, the coverage can then be approximated as

$$\theta = \frac{\frac{B}{T^{1/4}} e^{H_{Ads}/2k_B T}}{1 + \left(\frac{B}{T^{1/4}} e^{H_{Ads}/2k_B T}\right)}$$

When $H_{Ads}$ is negative, normal adsorption and desorption processes can occur, whereupon the coverage should decrease at higher temperatures. The resulting four parameters can then be varied to produce the fit line shown in Fig. 7, returning the physically meaningful values of $E_L = 1.3\,(.3)$ eV and $H_{Ads} = -1.6\,(.2)$ eV.

The removal of $O_2$ flow from the system at high temperature also causes the surface to recover back to reconstruction. The rate of recovery can also be modeled as a kinetic process in cases where there is a significant population of (1×1) in the initial state. Recovery rates are shown in the inset of Fig. 7. These rates are often faster than the rate of lifting at a given temperature, but also vary more slowly with temperature. The rate of (1×1) to hex can be fit to an Arrhenius relation, returning an activation barrier of 0.41 (0.14) eV indicated by $E_{hex}$ in Fig. 7b. This value is the same to within error as the earlier reported activation barrier for the recovery of the hex reconstruction after exposure to CO [31]. This correspondence is a further indication that this recovery process is limited by the rate of reconstruction and not that of oxygen desorption. In cases of very small (1×1) population, the rate of reconstruction forming showed less temperature dependence. The transient nature of (1×1) surface in the absence of flowing oxygen is an indication of a clean surface. In the instances of oxygen on Au (111) where no contaminants were observed the lifetime of the chemisorbed oxygen on the surface was often very short [4].

So far, we discussed only a simple first-order behavior of hex lifting and restoring kinetics. However, a more complicated kinetic can also occur at the higher temperatures (T > 1000 K) with the introduction of a sufficiently high oxygen flow. The lifting of the hex reconstruction would initially proceed as a simple first-order process with a rate constant consistent with the kinetics observed at lower flow rates. However, after a short period of time (10-30 seconds) from the onset of the first-order reaction, a much faster evolution process would take over the lifting process. A direct comparison of these two kinetic processes is shown in Fig. 8. These higher order processes with an incubation period typically begin to occur at pressures and temperatures just above the fit line in fully lifted region of the phase diagram. We may speculate that the lifted area slowly increases during the incubation period to a critical coverage at which point the lifting cascades cooperatively. We observed this behavior neither at low temperature nor at low $O_2$ flow rate.

In summary, there is a significant response to the Au (001) surface when exposed to molecular oxygen at high temperature. Relatively high pressures and temperatures over 800 K are required for this interaction to be observed and the resulting surface state is relatively unstable, quickly returning to a reconstructed Au (001) hex state once the oxygen flow is removed. While it was not possible within the scope of these experiments to determine the exact nucleation mechanism, the hex phase, normally rotated by 1 degree, lost the rotation and expanded upon exposure to oxygen indicating some degree of accommodation prior to lifting. Despite the well-known nobility of Au, this result is not entirely unsurprising as atomic oxygen has been known to interact with the Au (001) surface [16], as well as other reactive chemical species such as CO [31, 35, 36] and NO [37].

The exact nature of the oxygen adsorption, along with any potential dissociative mechanism at the surface will require theoretical guidance. DFT calculations indicate that adsorption lowers the energy relative to the molecular state [14]. This shift is seen as being responsible for the interaction, albeit one that produces adsorption only at high temperatures and oxygen pressures. It would be interesting to see if a molecular dynamics approach, such as that for the interaction of CO with the Pt (001) surface [38], would be capable of describing the behavior observed here. The observed kinetics here is an average of the surface processes and likely not comparable in a simplistic one-to-one fashion with theoretical calculations or the earlier TPD studies at lower temperatures [28]. There is also an open question as to whether or not the microstate of the surface is static or dynamic once equilibrium is reached during the coexistence regions. It should be possible to explore this possibility through the use of coherent surface x-ray scattering and x-ray photon correlation spectroscopy [39, 40, 41].

The collaboration is grateful for the valuable on-site assistance during the experiments provided by Dr. Evguenia Karapetrova and Dr. Christian Schlepuetz. This research used resources of the Advanced Photon Source, a U.S. Department of Energy (DOE) Office of Science User Facility operated for the DOE Office of Science by Argonne National Laboratory under Contract No. DE-AC02-06CH11357. The work at Safarik University received supported from European Regional Development Fund, Grant No. ITMS 26220120047. Data reduction and image processing was performed using the RIT Research Computing facilities. The portions of this work conducted at RIT were supported by the Research Corporation for Science Advancement (RCSA) through a Cottrell College Science Award.

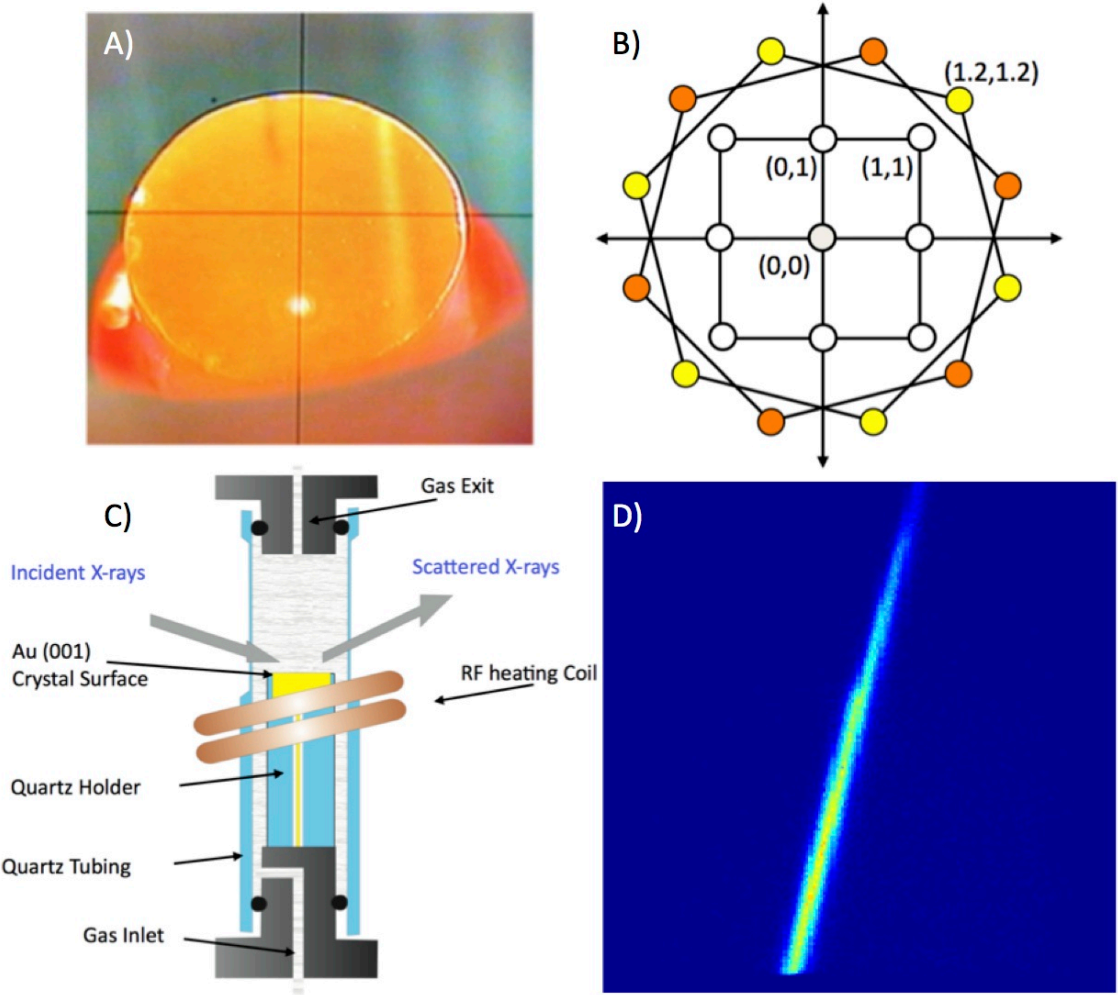

Figure 1. In A) at top left is an image of an Au (001) crystal at 1150 K. B) shows the surface reciprocal space diagram (L=0), demonstrating the 12 equivalent reflections for the two hex populations and the locations of the (0,H) and (H,H) surface rods. C) is a diagram of the experimental system., D) is an image of the hex peak taken at 1150 K.

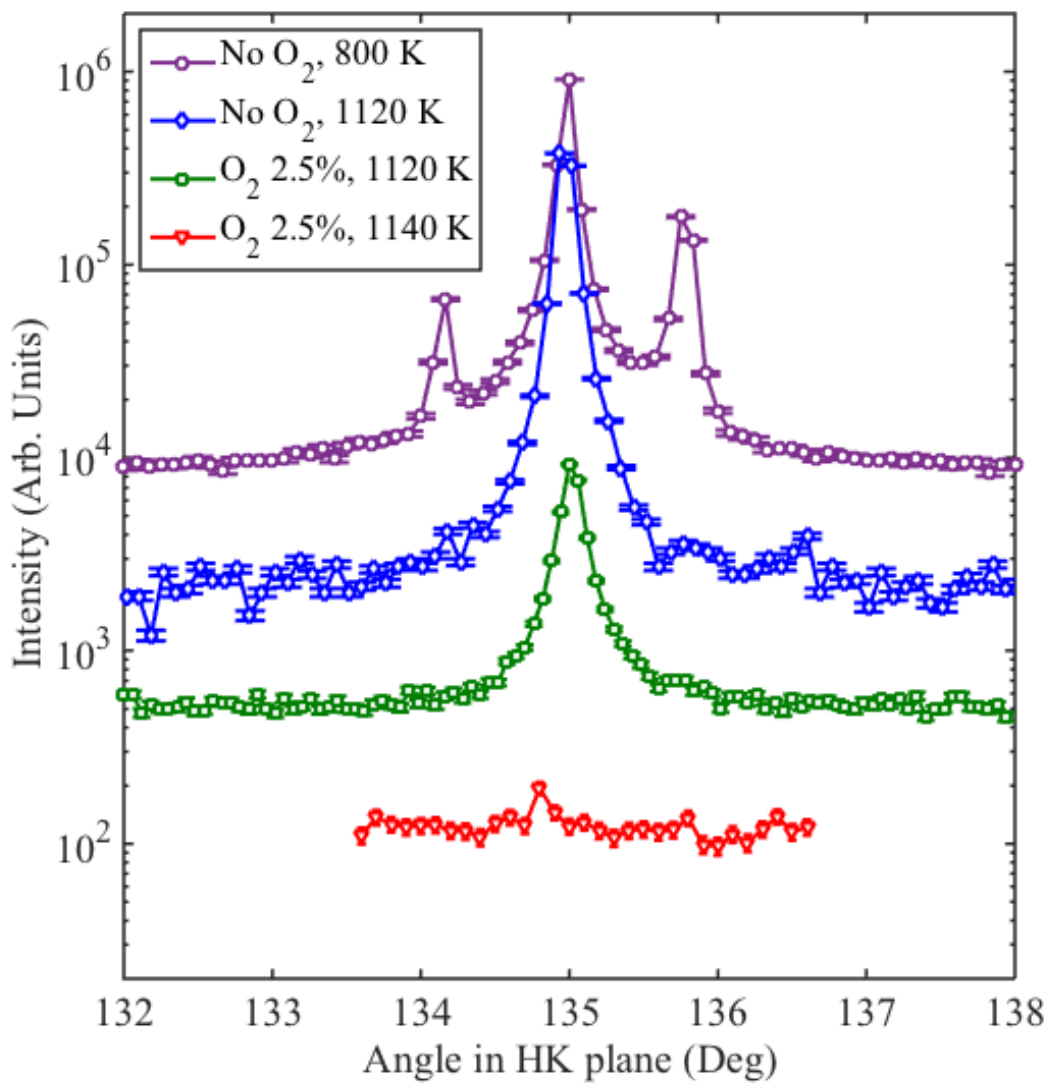

Figure 2. A log-scale plot of the intensity along a small in-plane arc near the hex peak in different conditions. Satellite peaks are present in the absence of oxygen at lower temperatures, while at higher temperatures the hex does not contain in-plane rotated phases. This is consistent with previous UHV studies of Au (001) [citation here]. If oxygen is added, the intensity of the hex peak is reduced or eliminated depending upon pressure and temperature. Each condition is scaled by a factor of 4 for clarity.

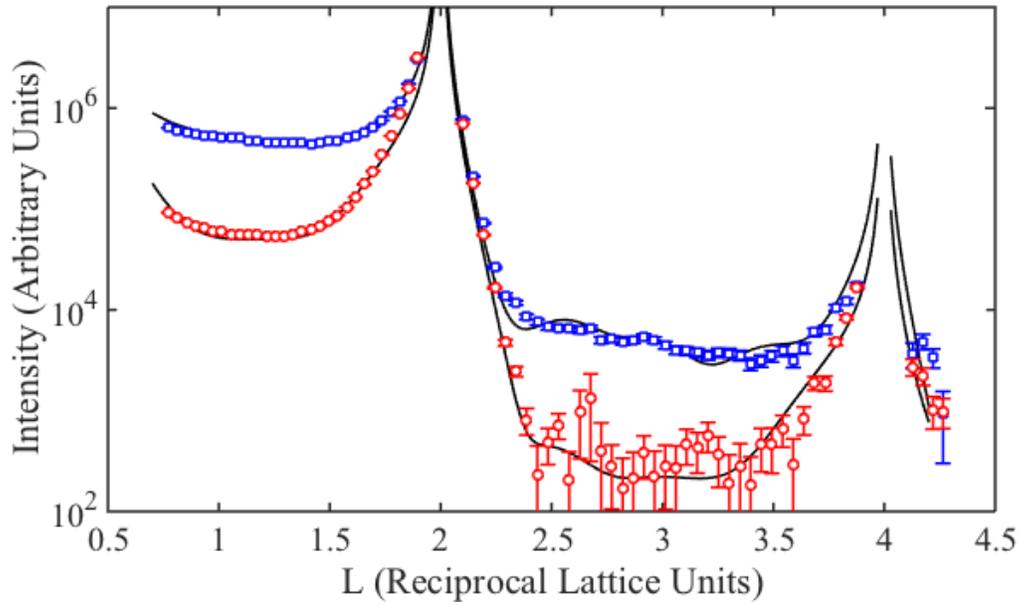

Figure 3. Specular crystal truncation rod data and fit lines for data collected at 1070 K, blue (squares) shows the absence of oxygen while red (circles) was taken with an oxygen partial pressure of 2.5%. The reduction in intensity due to oxygen is consistent with a disordered (1×1) terminated surface.

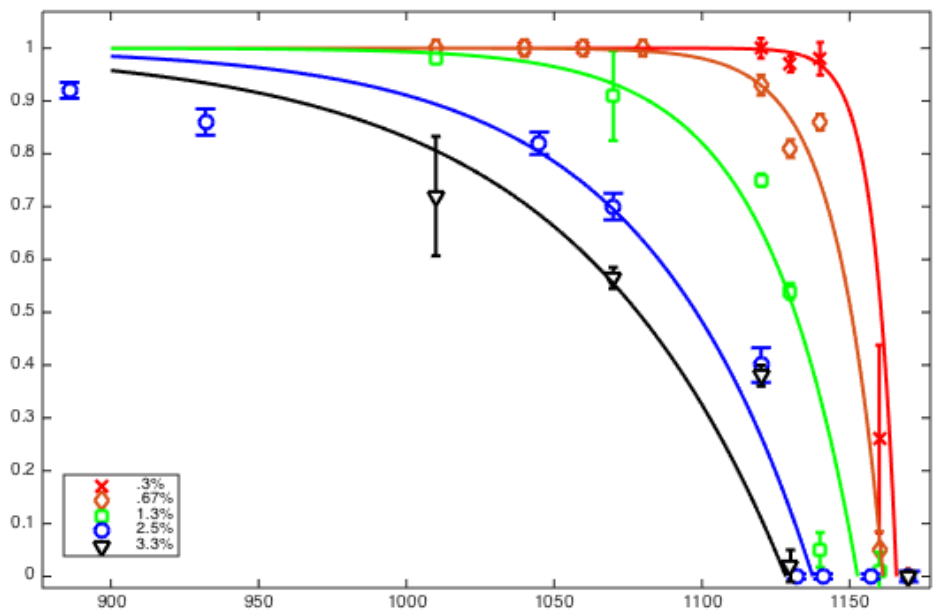

Figure 4. Isobar measurements of Au (001) surface reconstruction at different pressures, with corresponding fits lines from Equation 1.

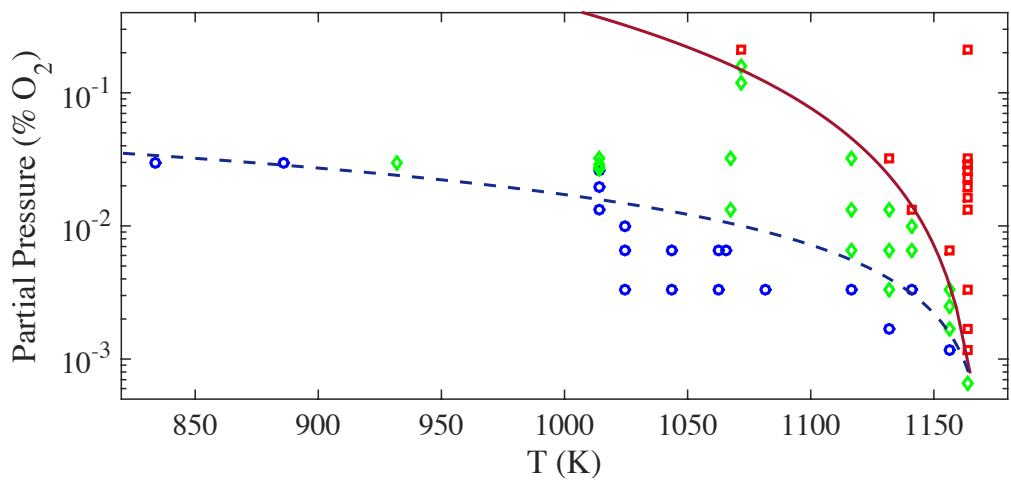

Figure 5. Pressure-Temperature phase diagram for the surface of the Au (001) – Oxygen system. At low pressures and temperatures ordered hex is observed with little or no change to the recorded intensity. There is an intermediate coexistence region where the x-ray intensity reduces, but does not vanish. At high pressures and temperatures the hex surface reconstruction lifts entirely, leaving only a disordered (1 × 1) surface.

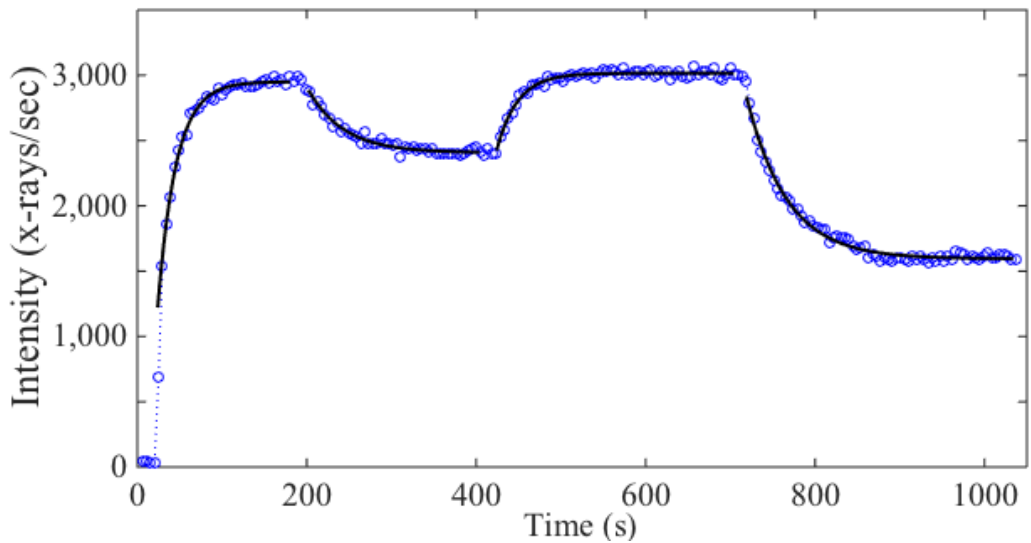

Figure 6. The dependence of the hex intensity peak at T = 1130 K upon exposure to oxygen. The oxygen flow rate began at 2.5% and was removed shortly after t=0 which results in the hex reconstruction appearing with full intensity. At t = 200 sec oxygen was reintroduced at a partial pressure of 0.5% atm and then later removed, followed by a partial pressure of 1.0% being added at t = 730 sec. The solid dark lines are fits using single exponentials and a constant. The rate of lifting did not depend upon the oxygen pressure in this example.

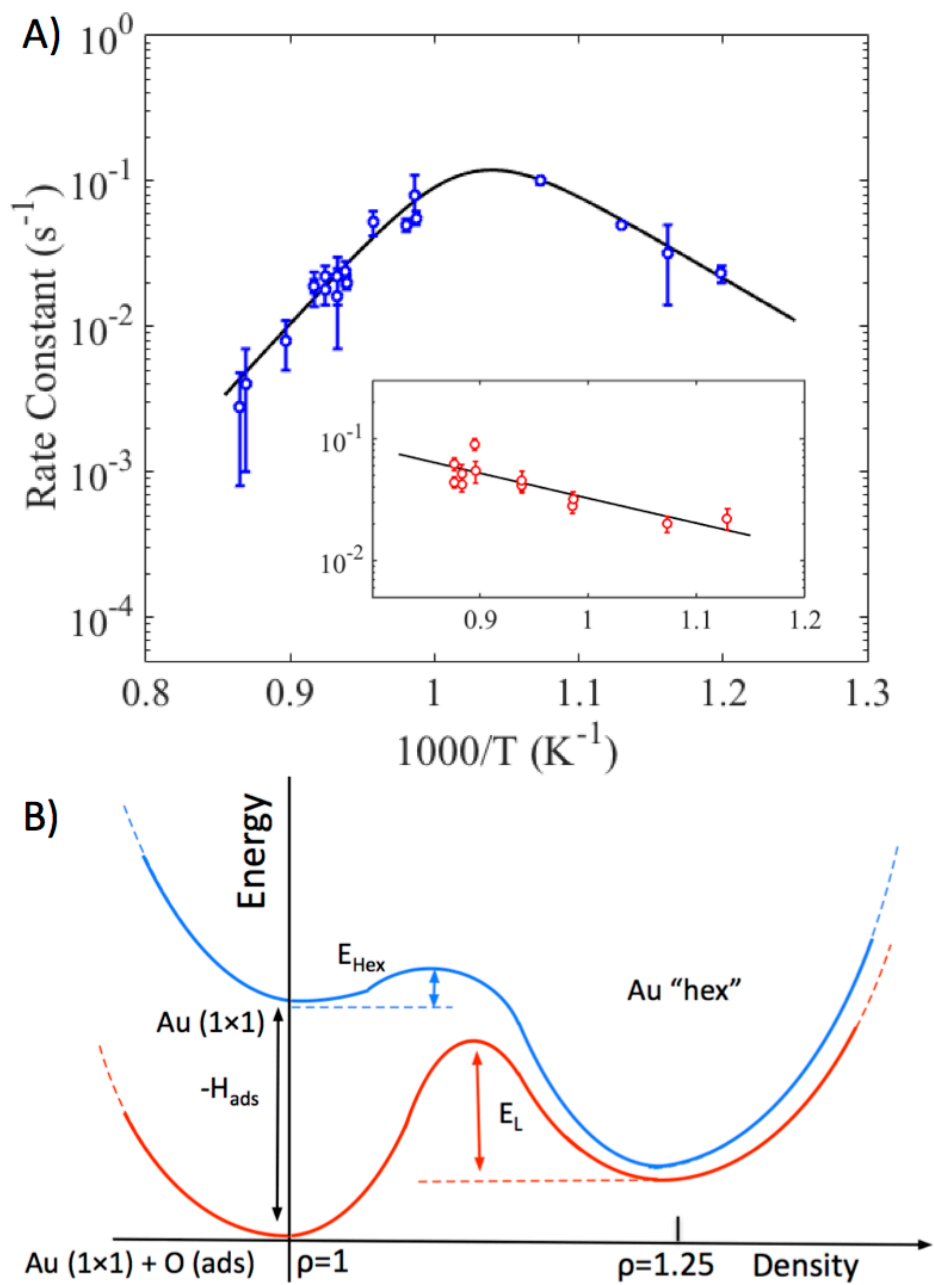

Figure 7. At top in A) the rate constants for hex to (1 × 1) plotted vs. inverse temperature using $O_2$ partial pressures in the range of 1% to 3%. The inset shows the rates for the recovery of the hex once $O_2$ is removed. At bottom, B) shows a simple energy diagram for the system.

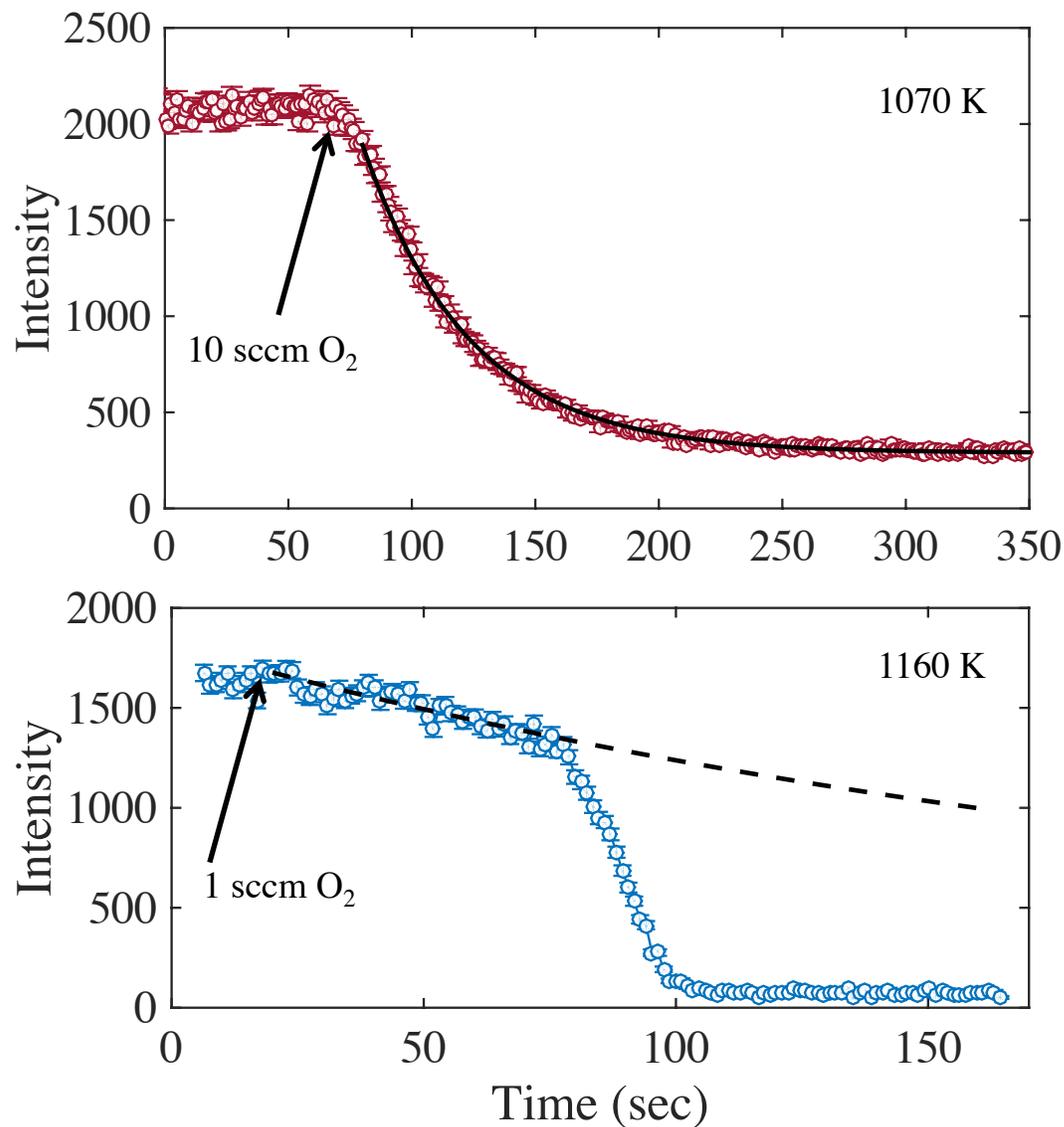

Figure 8. At top is an example of the simple first order kinetics observed at lower pressures and temperatures. The sample temperature was 1070 K and 10 sccm O2 was introduced to the cell at approximately t = 40 sec. The intensity of the hex peak then decays with a simple exponential fit. At bottom is an example of the higher pressure and temperature kinetics. The sample temperature was 1160 K and only 1 sccm of O2 was introduced at t = 20 sec. The intensity begins to slowly decay, with a simple exponential fit being shown, but at 80 sec the intensity then rapidly drops indicating that the process is no longer governed solely by first order kinetics.